\documentclass{aa}

\usepackage{graphicx,amsmath}
\usepackage{natbib}
\usepackage{bm}
\usepackage{float}

\usepackage{arydshln}

\usepackage{txfonts}
\usepackage{color}

\begin{document}

   \title{GALACTICNUCLEUS: A high-angular-resolution $JHK_s$ imaging survey of the Galactic centre}
   \titlerunning{GALACTICNUCLEUS}
   \authorrunning{Nogueras-Lara et al.}

   \subtitle{III. Evidence for wavelength dependence of the extinction curve in the near-infrared.}

  \author{F. Nogueras-Lara
          \inst{1}
          \and      
          R. Sch\"odel 
          \inst{2}       
          \and
          N. Neumayer 
          \inst{1}    
          \and
           E. Gallego-Cano
          \inst{2,3}
          \and
           B. Shahzamanian
          \inst{2}
         \and
          A. T. Gallego-Calvente
          \inst{2}    
          \and
           F. Najarro
          \inst{4}                        
          }

   \institute{
    Max-Planck Institute for Astronomy, K\"onigstuhl 17, 69117 Heidelberg, Germany
              \email{nogueras@mpia.de}
       \and 
           Instituto de Astrof\'isica de Andaluc\'ia (IAA-CSIC),
     Glorieta de la Astronom\'ia s/n, 18008 Granada, Spain                         
         \and
              Centro Astron\'omico Hispano-Alem\'an (CSIC-MPG), Observatorio Astron\'omico de Calar Alto, Sierra de los Filabres, 04550, G\'ergal, Almer\'ia, Spain     
         \and
              Centro de Astrobiolog\'ia (CSIC/INTA), ctra. de Ajalvir km. 4, 28850 Torrej\'on de Ardoz, Madrid, Spain                
}
   \date{}

  \abstract
   {The characterisation of the extinction curve in the near infrared (NIR) is fundamental to analyse the structure and stellar population of the Galactic centre (GC), whose analysis is hampered by the extreme interstellar extinction ($A_V\sim 30$ mag) that varies on arc-second scales. Recent studies indicate that the behaviour of the extinction curve might be more complex than previously assumed, pointing towards a variation of the extinction curve as a function of wavelength.}
   {We aim at analysing the variations of the extinction index, $\alpha$, with wavelength, line-of-sight, and absolute extinction, extending previous analysis to a larger area of the innermost regions of the Galaxy.}
   {We analysed the whole GALACTICNUCLEUS survey, a high-angular resolution ($\sim 0.2''$) $JHK_s$ NIR survey specially designed to observe the GC in unprecedented detail. It covers a region of $\sim 6000$\,pc$^2$, comprising fields in the nuclear stellar disc, the inner bulge, and the transition region between them. We applied two independent methods based on red clump (RC) stars to constrain the extinction curve and analysed its variation superseding previous studies.
}
{We used more than 165,000 RC stars and increased significantly the size of the regions analysed to confirm that the extinction curve varies with the wavelength. We estimated a difference $\Delta\alpha = 0.21\pm0.07$ between the obtained extinction indices, $\alpha_{JH}=2.44\pm0.05$ and $\alpha_{HK_s} = 2.23\pm0.05$. We also concluded that there is no significant variation of the extinction curve with wavelength, with the line-of-sight or the absolute extinction. Finally, we computed the ratios between extinctions, $A_J/A_H = 1.87\pm0.03$ and $A_{H}/A_{K_s} = 1.84\pm0.03$, consistent with all the regions of the GALACTICNUCLEUS catalogue.
}
   
   {}


\titlerunning{GALACTICNUCLEUS. III.}
\authorrunning{F. Nogueras-Lara et al.}

   \maketitle
%

 \section{Introduction}
 \label{intro}

The centre of the Milky Way is the closest galactic nucleus and the only one where we can resolve individual stars down to milli-parsec scales. It is, therefore, a unique laboratory to study the stellar nuclei and their role in the context of galaxy evolution. The Galactic centre (GC) is roughly delimited by the disc-like structure of the nuclear stellar disc (NSD) and the central molecular zone \citep[e.g.][]{Morris:1996vn,Launhardt:2002nx,Kruijssen:2014aa,Nogueras-Lara:2019ad}. It hosts a supermassive black hole, Sgr A*, in its dynamical centre that is embedded in a nuclear star cluster \citep[e.g.][]{Schodel:2014fk,Neumayer:2020aa}. 

The observation of the GC is hampered by the high stellar crowding and the extreme interstellar extinction ($A_V\gtrsim30$ mag, $A_{K_s}\gtrsim2.5$ mag, \citep[e.g.][]{Nishiyama:2008qa,Schodel:2010fk,Nogueras-Lara:2018aa}, that limits the stellar analysis to near/mid infrared observations. In this sense, the GALACTICNUCLEUS survey \citep{Nogueras-Lara:2018aa,Nogueras-Lara:2019aa} constitutes the most complete high-angular resolution ($\sim 0.2"$) catalogue available to properly characterise the stellar population in the GC. This is a near infrared (NIR) multi-wavelength ($JHK_s$) survey that covers a total area of $\sim0.3$ square degrees ($\sim6000$\,pc$^2$) along the NSD, the inner bulge, and the transition regions between the inner bulge and the NSD (see Fig. \ref{GNS}).

The characterisation of the extinction curve in the NIR is fundamental to fully exploit this data set and to be able to determine the structure and the stellar population of the innermost region of our Galaxy. Up to now, it has been widely accepted that it behaves like a power-law,  $A_{\lambda}\propto\lambda^{-\alpha}$ \citep[e.g.][]{Nishiyama:2008qa,Fritz:2011fk}, where $\lambda$ is the wavelength and $\alpha$, the extinction index. Nevertheless, recent studies have found some evidence of a wavelength dependence of the extinction index \citep{Nogueras-Lara:2018aa,Hosek:2018aa}. In particular, \citet{Nogueras-Lara:2019ac} analysed the extinction law in the NIR towards the central region of the NSD and found that the extinction index depends on wavelength, but not on the line-of-sight. This dependence has strong implications on the derivation of the structure of the innermost part of the Galaxy and the identification of the stellar type via NIR photometry, since a small change in the extinction index ($\sim 10-15\%$) leads to a significant change in absolute extinction \citep[$\sim0.3$ mag, e.g.][]{Matsunaga:2016aa,Nogueras-Lara:2019ac}, that might be translated into a biased estimation of the distance.

On the other hand, the study of the extinction curve using broadband filters is complex and requires to define a flux-weighted wavelength known as effective wavelength \citep[$\lambda_{eff}$, e.g.][]{Tokunaga:2005jw}. This quantity is not constant for a given filter and mainly depends on the absolute extinction and the spectral type of each star. The stellar metallicity, $log$g, and the atmospheric transmission also affect the value of $\lambda_{eff}$, but at a lesser level \citep[see Tables B.1 and B.2 in][]{Nogueras-Lara:2018aa}. The necessary use of the effective wavelength hampers the analysis of the extinction curve introducing degeneracies between different stellar types and the absolute value of extinction \citep[for further details, see appendix of][]{Nogueras-Lara:2018aa}. Namely, it is necessary to compute $\lambda_{eff}$ independently for each individual star to characterise the extinction curve. This requires to know the spectral type of each star before computing the extinction indices, which is impossible the high degeneracy existing  between differential extinction and stellar types in the NIR \citep[e.g.][]{Nogueras-Lara:2018aa}. To circumvent this problem, we use red clump stars (RC). They are red giants in their helium core burning sequence \citep[e.g.][]{Girardi:2016fk}, whose properties are well defined. They are abundant everywhere in the studied region and can be easily identified in the colour-magnitude diagrams (CMDs) using the NIR broadband filters $JHK_s$ (see Fig.\,\ref{CMD}). 

In addition, the large amount of differential extinction in the GC increases the complexity of the problem. We estimated that a variation in extinction of $\Delta K_s \sim 1$ mag (GC differential extinction estimated from Fig. \ref{CMD}), produces a change of the effective wavelength of RC stars of $\sim 0.5\,\%$ for $\lambda_{eff\_J}$ and $\lambda_{eff\_K_{s}}$, and $\sim 1\,\%$ for $\lambda_{eff\_H}$. These variations are apparently small, but result in differences of 0.05-0.13 for the estimated values of the extinction indices. This change leads to a wrong correction of the extinction, that makes the estimation of distance moduli or of stellar types from dereddening very difficult \citep[for further details see Figs. 33 and 34 of ][]{Nogueras-Lara:2018aa}. In this way, the uncertainties can be minimised by using RC stars and estimating the mean extinction towards regions of relatively small size.


In this work we analyse in detail the extinction curve towards the GC using $JHK_s$ photometry from the GALACTICNUCLEUS catalogue. We follow up the study initiated by \citet{Nogueras-Lara:2019ac} and extend it to the whole survey ($\sim 6000$ pc$^2$), by increasing the analysed area by a factor of 4. Our study includes regions that belong to the GC and to the inner bulge of the Galaxy.

 \section{Data}
 
\begin{figure}
   \includegraphics[width=\linewidth]{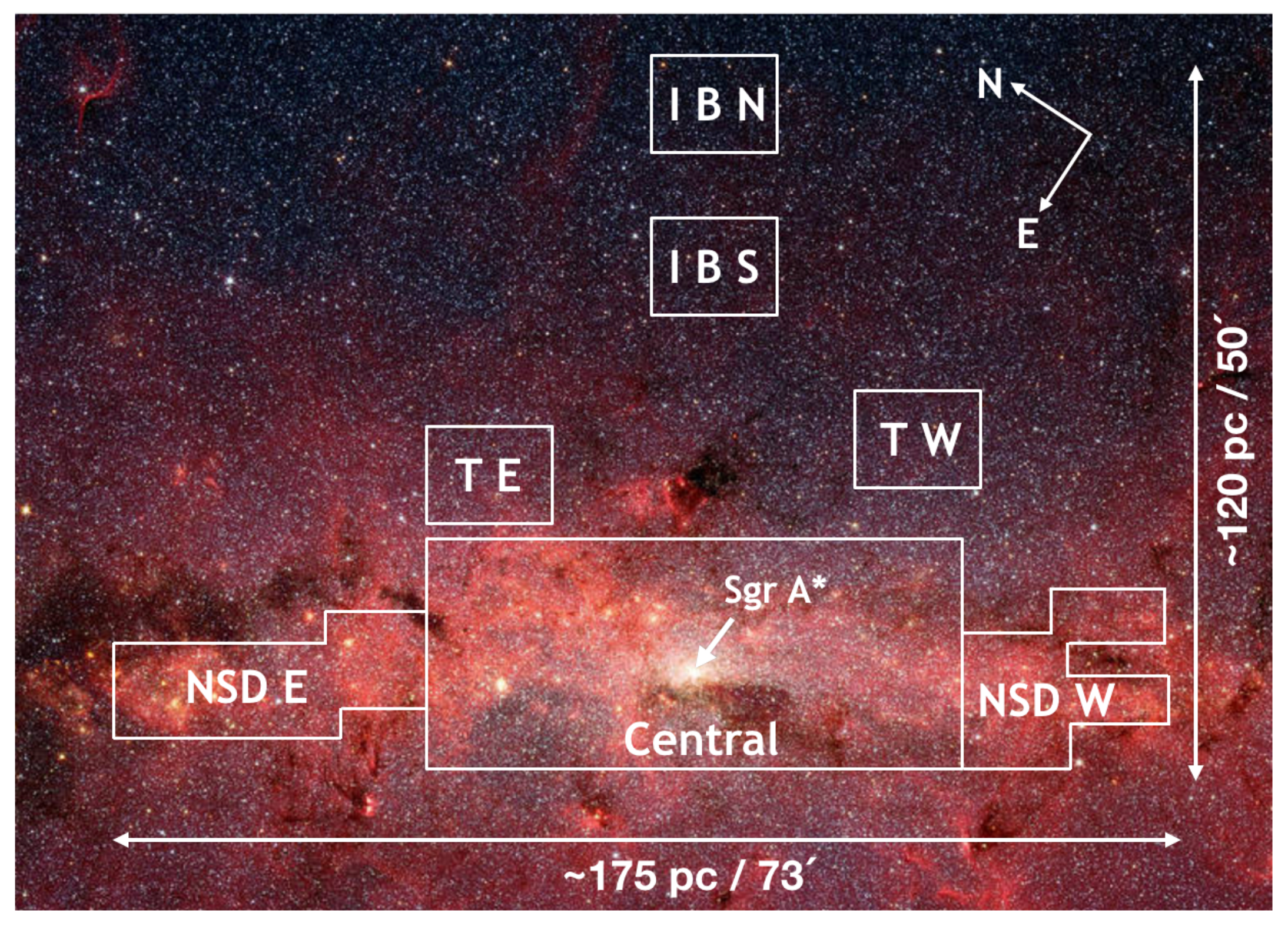}
   \caption{Scheme of the GALACTICNUCLEUS survey over-plotted on a false colour Spitzer/IRAC image at 3.6, 4.5, 5.8, and 8 $\mu$m (credits: NASA/JPL-Caltech/S. Stolovy (Spitzer Science Center/Caltech)). Each of the regions are indicated in the figure: Central, NSD East (NSD E), NSD West (NSD W), transition East (T E), transition West (T W), inner bulge South (I B S), and inner bulge North (I B N).}

   \label{GNS}
    \end{figure}

\begin{figure*}
   \includegraphics[width=\linewidth]{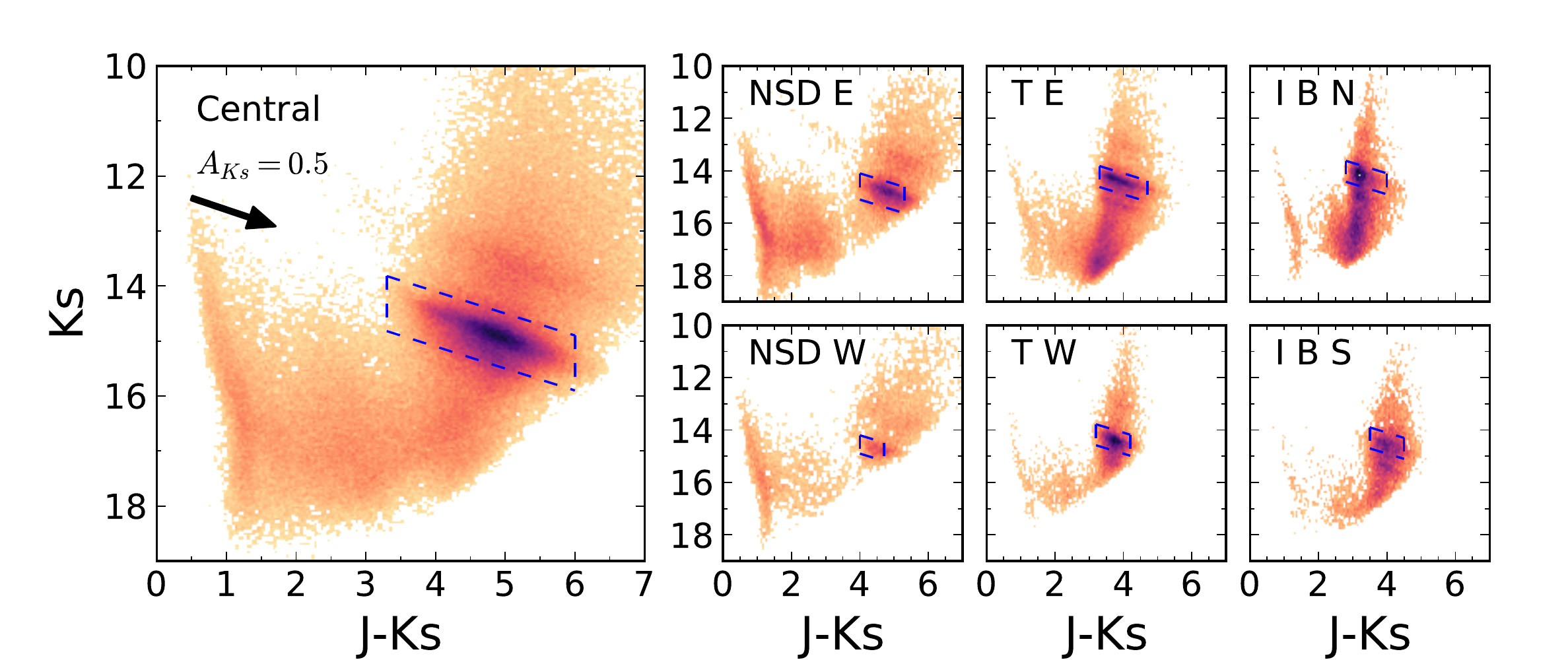}
   \caption{Colour-magnitude diagrams $K_s$ vs. $J-K_s$ of each of the regions observed by the GALACTICNUCLEUS survey: Central, NSD E, NSD W, TE, TW, IBN, and IBS correspond to the central region, the nuclear stellar disc East and West, the transition regions East and West, and the inner bulge North and South, as specified in Fig. 1. The blue dashed parallelograms indicate the position of the RC feature in each of the CMDs. The black arrow in the left panel depicts the reddening vector. The colour code indicates stellar densities, using a power stretch scale.}

   \label{CMD}
    \end{figure*}

 This work makes use of the GALACTICNUCLEUS survey \citep{Nogueras-Lara:2018aa,Nogueras-Lara:2019aa} carried out using the HAWK-I camera \citep{Kissler-Patig:2008uq} at the ESO VLT unit telescope 4, which includes accurate $JHK_s$ photometry of more than 3.3 million stars located in the GC and inner regions of the Galactic bulge. Figure \ref{GNS} shows the fields included in the GALACTICNUCLEUS survey: three different regions distributed along the GC (Central, NSD East, and NSD West), two regions in the inner Galactic bulge (inner bulge South and inner bulge North), and two transition regions (transition East and West). The photometry was measured with the {\it StarFinder} software \citep{Diolaiti:2000fk} that performs point spread function fitting and is optimised for crowded fields. The zero point (ZP) was calibrated using the SIRIUS IRSF survey \citep[e.g.][]{Nagayama:2003fk,Nishiyama:2006tx} and has a systematic uncertainty of 0.04 mag for all three bands. The photometry reaches 5\,$\sigma$ detections for $J\sim22$, $H\sim21$, and $K_s\sim21$ mag. The statistical uncertainties are below 0.05 mag at $J\lesssim21$, $H\lesssim19$, and $K_s\lesssim18$ \citep{Nogueras-Lara:2019aa}. This allows us to properly cover RC stars \citep[e.g.][]{Girardi:2016fk} at the GC distance and extinction conditions. Figure \ref{CMD} shows the colour-magnitude diagram $K_s$ versus $J-K_s$ of all the regions covered by the GALACTICNUCLEUS survey. The blue dashed parallelogram indicates the RC feature in each of the panels.

\section{Extinction index variability with the wavelength}
\label{EIVW}

We used two different methods to compute the extinction indices between $JH$ and $HK_s$.

\subsection{Slopes of the RC feature}
\label{slope_sec}

The slope of the RC feature in the CMDs indicates the direction of the reddening vector and can be used to compute the extinction index following Eq. (1) in \citet{Nogueras-Lara:2018ab, Nogueras-Lara:2019ac}:

\begin{equation}
\label{eq_slope}
\alpha = -\frac{\log(1+\frac{1}{m})}{\log(\frac{\lambda_{\text{eff}_1}}{\lambda_{\text{eff}_2}})},
\end{equation}

\noindent where $\alpha$ is the extinction index, $\lambda_{eff_1}$ and $\lambda_{eff_2}$ are the effective wavelengths of the considered bands, and $m$ is the slope of the RC feature in the CMD.

To apply this method: (1) The differential extinction must be large enough to result in an RC feature extended enough to allow us a reliable measurement of its slope. (2) It is necessary to have some previous information about the star formation history (SFH) of the analysed region. In particular, the brightness of RC stars depends on their ages, metallicity, and/or enhancement in alpha elements \citep[e.g.][]{Girardi:2016fk, Nogueras-Lara:2018ab}. Moreover, a secondary clump formed by the red giant branch bump (RGBB) might also appear depending on the age and metallicity of the stellar population. Due to variations in the distance and line-of-sight extinction, the RC and RGBB features can be blended. Also, there can be degeneracies between the RGBB and a possibly fainter RC feature from a $\sim$1\,Gyr old burst of star formation.

In the subsequent analysis, we only used the central region of the GALACTICNUCLEUS survey and the nuclear stellar disc East (NSD E, see Fig. \ref{CMD}). This is due to the required quality and the previous knowledge on the SFH \citep{Nogueras-Lara:2019ad}. Moreover only in these fields, the differential extinction is large enough to reliably measure the slope of the RC feature in the CMDs. We excluded the nuclear stellar disc West due to the bad quality of the data (too few stars detected due to bad data quality caused by bad weather conditions). The transition regions (T E and T W) were not considered due to the mixture of populations between the inner bulge and the NSD that can affect the calculation of the slopes. We also excluded the inner bulge regions given that the differential extinction in the $K_s$ versus $H-K_s$ diagram is much smaller than in the central regions, and this effect might lead to an incorrect estimation of the slopes.

Recent work on the NSD by \citet{Nogueras-Lara:2019ad}, points towards a mainly old stellar population ($>80\,\%$ of the mass older than 8 Gyr) and an important star formation event $\sim$\,1\,Gyr ago ($>5\,\%$ of the mass), implying a double RC sequence in the CMDs \citep[see Fig. 1 of ][]{Nogueras-Lara:2019ad}.  \citet{Nogueras-Lara:2019ac} analysed in detail the central $\sim$\,1700 pc$^2$ of the GALACTICNUCLEUS survey. Here, we increase the studied area  by a factor $\sim 2$, and also analyse the NSD E field. 

\begin{figure}
   \includegraphics[width=\linewidth]{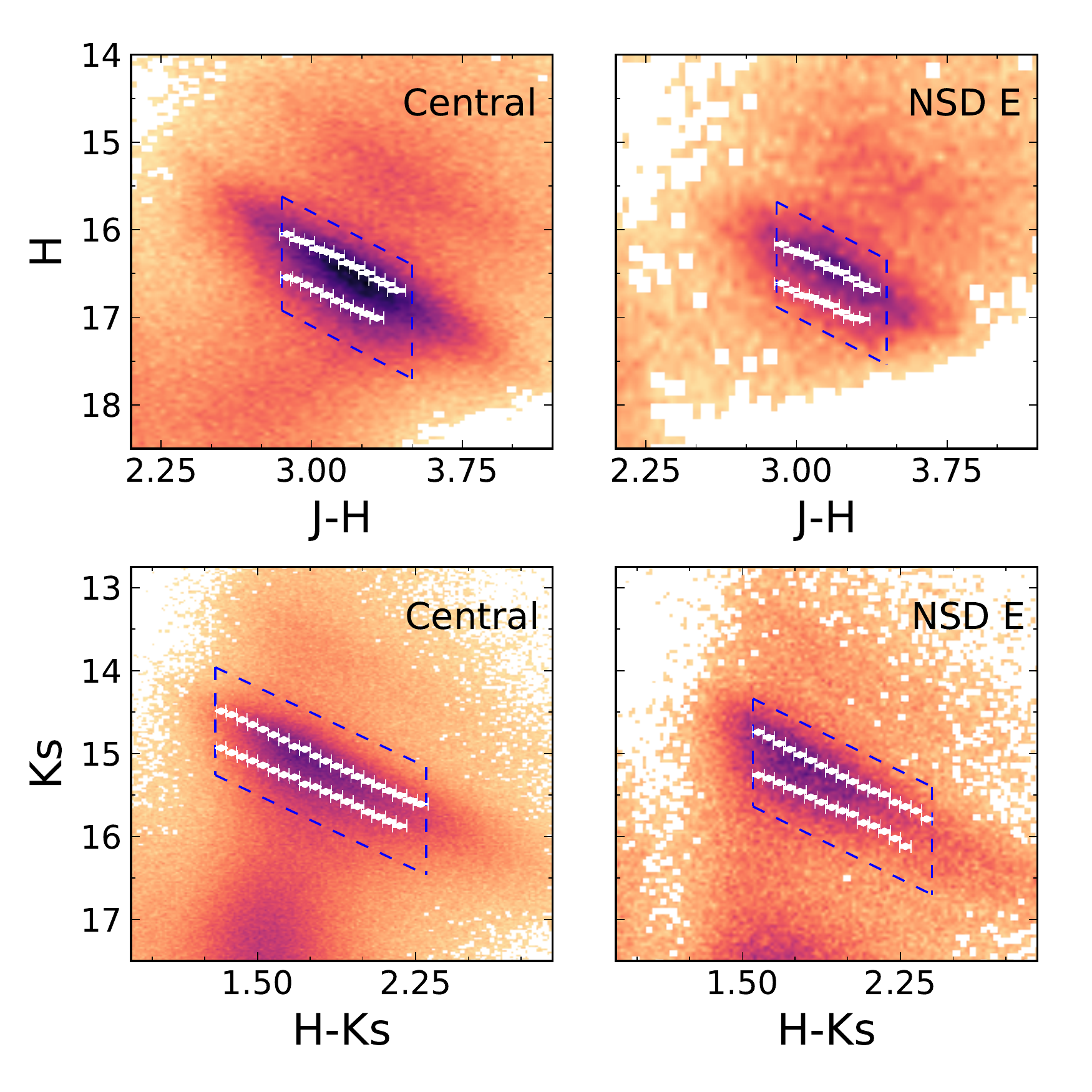}
   \caption{Colour-magnitude diagrams $H$ vs. $J-H$ of (upper panels), and $K_s$ vs. $H-K_s$ (lower panels) for the central region of the GALACTICNUCLEUS survey (left panels), and the NSD East (right panels). The blue dashed parallelograms indicate the stars considered to compute the slope of the RC features. White dots depict the solutions obtained when applying the GMM method and their associated uncertainties.}

   \label{slope}
    \end{figure}

We used all the stars in the RC features as shown by the blue dashed parallelograms in Fig. \ref{slope}. The CMDs include stars belonging to several different pointings from the GALACTICNUCLEUS survey that were obtained under different observing conditions and different dates \citep[see Tables A1-A3 in][]{Nogueras-Lara:2019aa}. Moreover, the differential extinction varies significantly across the observed field \citep{Nogueras-Lara:2018aa,Nogueras-Lara:2019ac}, which influences the position of the RC features in the CMDs depending on the line-of-sight. Thus, the selection of the RC is made in a way that the RC features present a homogeneous density in the CMDs and are not significantly affected by foreground population and/or low completeness. In this way, we excluded the bright part of the RC feature, since it might contain stars belonging to the inner bulge \citep{Nogueras-Lara:2019ad}, and the faint end, since it is more affected by extinction and incompleteness.

To analyse the RC features and compute their slopes, we applied the methodology described in Sect. 3 of \citet{Nogueras-Lara:2018ab}, and Sect. 4.4 of \citet{Nogueras-Lara:2019ac}. Namely, we divided the RC regions into small vertical bins of 0.05 mag width and confirmed that they correspond to a double RC sequence. For this, we considered each vertical bin and fitted the underlying $K_s$-band distribution with a one-Gaussian model and a two-Gaussians one, applying the SCIKIT-LEARN python function GaussianMixture \citep[GMM][]{Pedregosa:2011aa}. We obtained that a two-Gaussians model fits the data better in all cases, using the Bayesian information criterion  \citep{Schwarz:1978aa} and the Akaike information criterion \citep{Akaike:1974aa}, as expected given the SFH of the analysed regions. To compute the slope of the features, we used a jackknife algorithm considering the values obtained from the two-Gaussians fits and the median colours of the vertical bins (see Fig. \ref{slope}). The slopes were obtained as the mean of the re-sampling data sets in the jackknife algorithm and the associated uncertainties were estimated as their variances. We calculated the extinction indices using Eq. \ref{eq_slope}. The effective wavelengths were computed following the Eq. (A3) of \citet{Tokunaga:2005jw}, as explained in Appendix B of \citet{Nogueras-Lara:2018aa}, using an index of $2.30\pm0.08$ \citet{Nogueras-Lara:2018aa} and the $A_{1.61}$ values shown in Table\,\ref{alpha_grid}.

Table \ref{alpha_slope} shows the results, where the statistical and systematic uncertainties are specified. The statistical uncertainties were estimated considering the uncertainties of the slopes and the effective wavelengths \citep[see appendix B of ][]{Nogueras-Lara:2018aa}. To obtain the systematics, we varied the width and the number of vertical bins (implying different cuts at the faint end), and the selection box of the RC stars (blue dashed parallelograms in Fig. \ref{slope}). In all cases, we removed the faint red end of the secondary RC feature to avoid problems related to the higher incompleteness of this feature \citep{Nogueras-Lara:2019ac}. The extinction indices obtained using the faint features are somewhat larger than the ones obtained by means of the bright one. We believe that the reason may be that the secondary RC is fainter and has a significantly lower number of stars than the bright one. Therefore, it is significantly more affected by incompleteness, differential extinction, and the presence of recent star formation. In this way, we calculated the extinction indices combining the obtained values for both features to account for possible systematic effects. Table \ref{alpha_slope} shows the average values and their associated uncertainties.

\subsubsection{Wavelength variability}

Comparing the extinction indices computed previously for $JH$ and $HK_s$ ($\Delta\alpha=\alpha_{JH}-\alpha_{HK_s}$), we obtained $\Delta\alpha_{central} = 0.19\pm0.05$ and $\Delta\alpha_{NSD\ E} = 0.34\pm0.06$, where the statistical and systematic uncertainties of each $\alpha$ were propagated quadratically. Combining both measurements, we obtained $\Delta\alpha=0.27\pm0.04$, where the uncertainty was computed propagating the uncertainty of each measurement. This implies that the difference in the extinction indices is detected with $\gtrsim 6\  \sigma$ significance.

\begin{table}
\caption{Extinction index calculation following the method described in Sect. \ref{slope_sec}.}
\label{alpha_slope} 
\begin{center}
\def\arraystretch{1.2}
\begin{tabular}{ccc}
 &  & \tabularnewline
\hline 
\hline 
Extinction index & Central & NSD E\tabularnewline
\hline 
$\alpha_{JH\_bright} $& $2.43\pm0.05\pm0.03$ & $2.43\pm0.04\pm0.03$\tabularnewline
$\alpha_{JH\_faint} $& $2.53\pm0.05\pm0.06$ & $2.61\pm0.10\pm0.09$\tabularnewline
$\alpha_{JH\_average} $& $2.48\pm0.04\pm0.03$ & $2.52\pm0.05\pm0.05$\tabularnewline
\hline 
$\alpha_{HK_s\_bright} $& $2.21\pm0.01\pm0.02$ &$ 2.13\pm0.02\pm0.01$\tabularnewline
$\alpha_{HK_s\_faint} $& $2.36\pm0.01\pm0.03$ & $2.23\pm0.05\pm0.03$\tabularnewline
$\alpha_{HK_s\_average} $& $2.29\pm0.01\pm0.02$ & $2.18\pm0.03\pm0.02$\tabularnewline
\hline 
\end{tabular}

\end{center}
\vspace{0.5cm}
\textbf{Notes.} The uncertainties correspond to the statistical and systematic ones, respectively.

 \end{table}

\subsubsection{Completeness effect}
\label{comple}

We also analysed the effect of completeness on our results. Given the high number of sources in our fields, the standard approach of inserting and recovering artificial stars is not feasible due to enormous amount of computational time required. Instead, we used an alternative approach computing the critical distance at which a star of any magnitude might be detected around a given brighter star \citep{Eisenhauer:1998tg}. This allows us to account for completeness due to crowding, which dominates the incompleteness in the highly crowded regions studied \citep{Nogueras-Lara:2019ad}. We considered the central region of the GALACTICNUCLEUS survey as a test case. This is because this is the most crowded region of the survey and the effect of incompleteness due to crowding will be maximum here. Thus, if there is some influence of incompleteness on our results, it will be easily identified in this region.  Since this method assumes that the probability of detecting a source with a given magnitude is uniform within the field, we divided the region into small sub-regions of $2' \times 1.4'$. We averaged over the completeness solutions obtained for each sub-region to get the final one \citep[for further details, see][]{Nogueras-Lara:2019ad}. We ended up with 80\,\% of completeness level for $J\sim18.4$ mag, $H\sim18.3$ mag, and $K_s\sim16.3$ mag. Nevertheless, the completeness for $J$ band (and presumably $H$ band as well) is probably overestimated since the crowding becomes the limiting factor for longer wavelengths (i.e. $K_s$ band). This is due to two main reasons: (1) The shorter the wavelength, the more important is the extinction effect and less stars are detected. Thus, the incompleteness due to the lack of sensitivity is significant and the crowding effect is less important. (2) The luminosity function is very similar to a power law outside of the RC (and at $J$ band the differential extinction dilutes the RC extremely, so that a power-law is a good approximation). Therefore, if the turnoff of the LF is at brighter magnitudes than the crowding completeness, then completeness is limited by sensitivity.

We corrected the CMDs for completeness choosing a reference level with a completeness of 50\,\%, and randomly removing stars from the CMDs  whose completeness is larger to normalise them to the reference level. For this, we computed the completeness levels on the CMDs $H$ vs. $J-H$ and $K_s$ vs. $H-K_s$ in steps of 1\,\% to calculate the random fraction of stars to be removed in each 1\,\% step. We generated 100 Monte Carlo samples of randomly removed stars and repeated the calculation of the extinction index using the bright RC feature of the CMDs, as it was explained previously. We obtained that the standard deviation of the values is $<0.01$ for both $\alpha_{JH}$ and $\alpha_{HK_s}$. Therefore, we concluded that the effect of completeness is negligible given the chosen selection of RC stars (blue parallelograms in Fig. \ref{slope}).

We also checked the influence of completeness using different brightness cutoffs when computing the slopes of the features. We concluded that, within the uncertainties and the selection of RC stars used here, there is not any significant effect of completeness on our results.

\subsubsection{Effect of the nuclear star cluster}

The central field of the GALACTICNUCLEUS survey includes the nuclear star cluster (NSC), whose SFH might be different in comparison to the NSD. In addition, the significantly higher stellar density might result in a lower completeness of the data that might affect the RC features shown in Fig. \ref{slope}. Namely, according to previous works, around 70-80\,\% of the stars are older than 5 Gyr \citep{Blum:2003fk,Pfuhl:2011uq}. Therefore, the bright RC feature should not be significantly affected. Nevertheless, and as a sanity check, we repeated the analysis excluding all the stars within a radius of 5 pc ($\sim$ 2.2 arcmin) from Sgr\,A*, that corresponds to the effective radius of the NSC \citep[e.g.][]{gallego-cano2019}. Using the bright RC feature we obtained $\alpha_{JH}=2.40\pm0.05\pm0.03$ and $\alpha_{HK_s}=2.20\pm0.01\pm0.02$, in good agreement with the results obtained previously. Comparing the extinction indices we computed $\Delta\alpha = 0.20\pm0.06$, where the uncertainties have been added quadratically.

\subsection{Model minimisation}
\label{grid}

We also analysed the extinction curve using RC stars and the {\it grid method} described in \citet{Nogueras-Lara:2018aa}. This method assumes a Kurucz  model \citep{Kurucz:1993fk} for a RC star and generates a grid of extinction indices ($\alpha$) and absolute extinctions ($A_{1.61}$) at a given wavelength ($\lambda={1.61}\,\mu$m). We made the grid finer with step sizes $\sim 3$ times smaller compared to previous work (step of 0.005 for $\alpha$ and $A_{1.61}$). Applying the grid, we reddened the synthetic model to be compared with the real data via $\chi^2$ minimisation. The RC stellar model considers T=$4750\pm250$\,K, log $g$=+2.5 \citep{2014ApJ...790..127B}, a radius of $10.0\pm 0.5$\,$R_\odot$ \citep[e.g.][]{Chaplin:2013kx}, and twice solar metallicity \citep[e.g.][]{Feldmeier-Krause:2017kq, Schultheis:2019aa, Nogueras-Lara:2019ad}. We assumed a distance to the GC of $8.0\pm0.1$ kpc, combining the results obtained by \citet{Gravity-Collaboration:2018aa,Do:2019aa}. To estimate the systematic uncertainties, we repeated the calculation of $\alpha$ and $A_{1.61}$ considering independently the uncertainties of each of the parameters described previously \citep[for further details see][]{Nogueras-Lara:2018aa, Nogueras-Lara:2019ac}. Given the high number of stars used and the small step size of the grid, the computational time to estimate the uncertainties is too high. Therefore, we randomly selected a sample of 5000 stars for each of the GALACTICNUCLEUS' regions (we selected all the stars for regions where the number of RC stars is lower than 5000) to compute the systematics. Moreover, we also considered the systematic uncertainty of the photometric zero point, repeating the calculation of $\alpha$ and $A_{1.61}$ adding and subtracting the ZP systematics to the photometry of each band independently \citep[0.04 mag in all three bands, ][]{Nogueras-Lara:2019aa}. The statistical uncertainties are not relevant in any case given the large number of stars used for the calculations.

We applied this method to the RC stars in the whole GALACTICNUCLEUS survey to study the variability of the extinction curve. Since we used only two bands to compute two unknowns ($\alpha$ and $A_{1.61}$), the grid method is similar to a geometric method in the CMD space, implying that  might be dependent on the stellar density of the selected RC features. Thus, to avoid selection effects, we used the CMD $K_s$ vs. $J-K_s$ when analysing the extinction indices for $JH$ and $HK_s$. The blue dashed parallelograms in Fig. \ref{CMD} indicate the RC stars used for each region. This selection depends on the stellar density of the RC feature for each field. In this way, the RC feature is narrower for the transition and the inner bulge fields than in the regions belonging to the NSD. This is because the RC there is mainly old and does not have much contribution from stellar populations younger than 8\,Gyr \citep{Nogueras-Lara:2018ab,Nogueras-Lara:2019ad}. Moreover, we only used stars detected in all three bands and excluded the faint end of the RC feature to avoid regions with low completeness.

\begin{figure}
   \includegraphics[width=\linewidth]{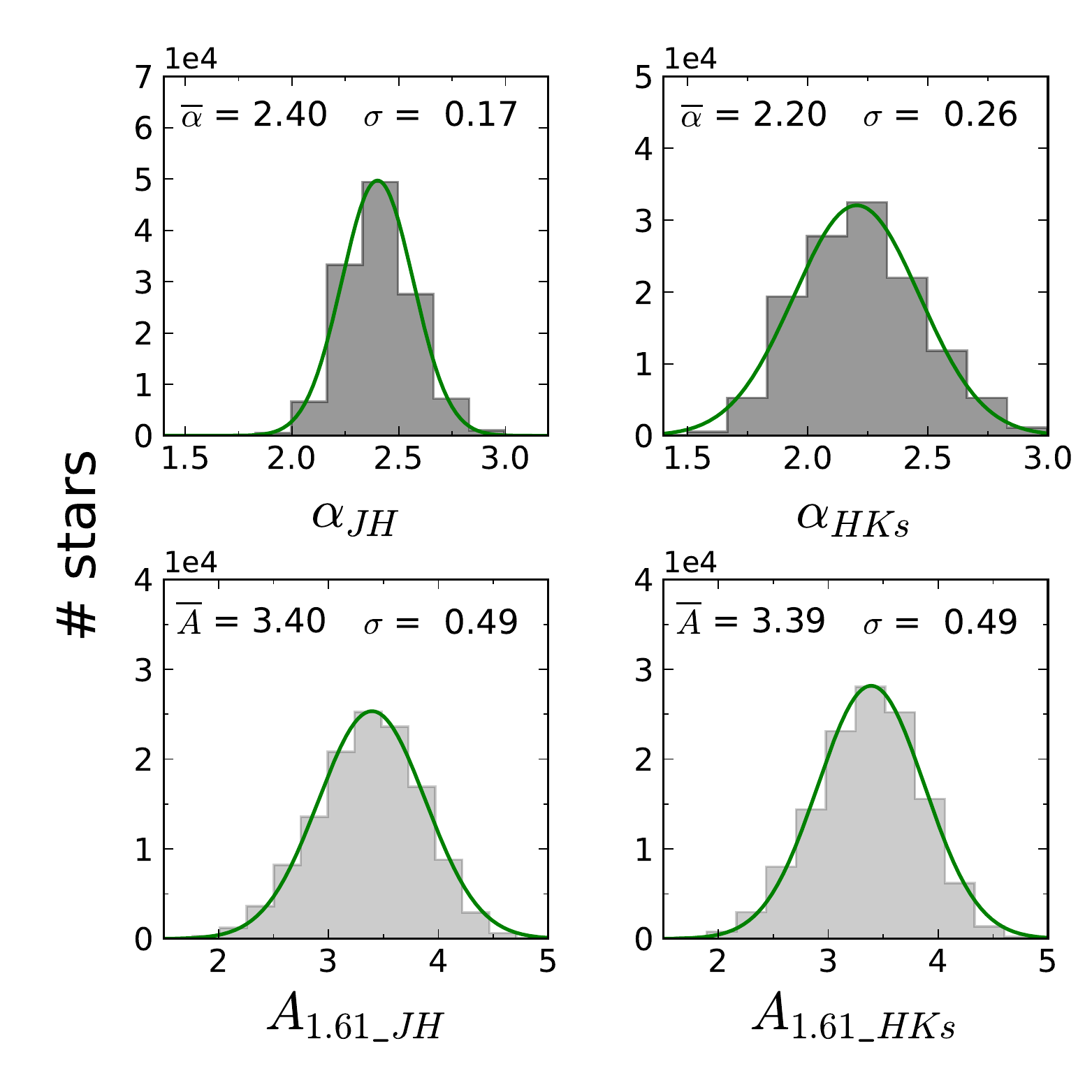}
   \caption{Extinction indices ($\alpha_{JH}$ and $\alpha_{HK_s}$) and absolute extinction ($A_{1.61\_JH}$ and $A_{1.61\_HK_s}$) distributions, upper and lower panels, respectively. The green line indicates a Gaussian fit whose mean and standard deviations are specified in each panel.}

   \label{grid_figure}
    \end{figure}

Figure \ref{grid_figure} shows the obtained $\alpha$ and $A_{1.61}$ distributions for the central region of the GALACTICNUCLEUS survey, as an example. Table \ref{alpha_grid} summarises the results obtained for each region in the catalogue.  We found that the $\alpha_{HK_s}$ uncertainties are a factor $\sim 2$  larger than the ones for $\alpha_{JH}$. This is because variations in the ZP, the distance to the stars, and/or the temperature and radius of the RC stars affect the calculation of $\alpha_{HK_s}$ in a more significant way. On the other hand, to assess our results, we confirmed that  the values of extinction, $A_{1.61\_JH}$ and  $A_{1.61\_HK_s}$, obtained when computing the extinction indices using the bands $JH$ and $HK_s$, agree, as expected (Table \ref{alpha_grid}).

\begin{table*}
\caption{Obtained values for $\alpha$ and $A_{1.61}$ using the grid method.}
\label{alpha_grid} 
\begin{center}
\def\arraystretch{1.4}
\setlength{\tabcolsep}{3.8pt}
\begin{tabular}{cccccc}
 &  &  &  & & \tabularnewline
\hline 
\hline 
 & $\alpha_{JH}$ & $\alpha_{HK}$ & $A_{1.61\_JH}$ & $A_{1.61\_HK}$ & $\Delta\alpha$\tabularnewline
\hline 
Central & $2.40 \pm 0.07$ & $2.20 \pm 0.12$ & $3.40 \pm 0.12$ & $3.39 \pm 0.12$ & $0.20 \pm 0.12$\tabularnewline
NSD E & $2.42 \pm 0.07$ & $2.25 \pm 0.12$ & $3.29 \pm 0.12$ & $3.29 \pm 0.12$ & $0.17 \pm 0.12$\tabularnewline
NSD W & $2.38 \pm 0.08$ & - & $2.96 \pm 0.12$ & -  & -\tabularnewline
T E & $2.45 \pm 0.09$ & $2.32 \pm 0.16$ & $2.56 \pm 0.12$ & $2.56 \pm 0.12$ & $0.13 \pm 0.15$\tabularnewline
T W & $2.45 \pm 0.10$ & $2.21 \pm 0.16$ & $2.47 \pm 0.12$ & $2.47 \pm 0.12$ & $0.24 \pm 0.15$\tabularnewline
I B S & $2.40 \pm 0.08$ & $2.21 \pm 0.15$ & $2.71 \pm 0.12$ & $2.70 \pm 0.12$ & $0.19 \pm 0.14$\tabularnewline
I B N & $2.50 \pm 0.11$ & $2.17 \pm 0.18$ & $2.12 \pm 0.12$ & $2.11 \pm 0.12$ & $0.33 \pm 0.17$\tabularnewline
\hline 
\end{tabular}

\end{center}
\textbf{Notes.} $A_{1.61\_JH}$ and $A_{1.61\_HK}$ indicate the value of the extinction at 1.61 $\mu$m obtained using the bands $JH$ and $HK_s$, respectively. The uncertainties correspond to the systematic ones. The statistical uncertainties are negligible given the high number of stars used for the calculation. The $\alpha_{HK}$ and $A_{1.61\_HK}$ values for the NSD W were not computed due to the low number of stars detected in the RC feature and the bad weather conditions when observing $K_s$ band. It affected the photometry as indicated by the scattering in Fig. 7 of \citet{Nogueras-Lara:2019aa} and might lead to biased results.

 \end{table*}

\subsubsection{Wavelength variability}

We computed $\Delta\alpha$ for each region using the values obtained for $\alpha_{JH}$ and $\alpha_{HK_s}$.  We estimated the uncertainties repeating the calculation of  $\alpha_{JH}$ and $\alpha_{HK_s}$ considering the uncertainties of the parameters involved in their calculation. We used again the approach of selecting a random sample of 5000 RC stars to avoid too high computational times (see Sect. \ref{grid} for details). For each of the parameters, we calculated the uncertainty on $\Delta \alpha$ and propagated them quadratically. Since the variation of some of the parameters affects $\alpha_{JH}$ and $\alpha_{HK_s}$ in the same direction, the final uncertainty is lower than the obtained when propagating the individual uncertainties computed previously for each extinction index (Table \ref{alpha_grid}). The $H$ and $K_s$ ZP systematics are the most important source of uncertainty for $\Delta \alpha$. The last column of Table \ref{alpha_grid} shows the results. We observed that the extinction indices, $\alpha_{JH}$ and $\alpha_{HK_s}$, are different for all the regions analysed with $\sim 2 \sigma$ significance. Computing the $\Delta \alpha$ averaging over the values obtained for each of the regions, we ended up with a value of $\Delta \alpha = 0.21 \pm 0.06$, where the uncertainty refers to the standard deviation of the  distribution of $\Delta \alpha$.

\subsubsection{Completeness effect}

We also estimated the influence of completeness on our results. We considered the central region and followed the approach described previously (see Sect. \ref{comple}). We computed the completeness solution for $J$ and $K_s$, since we used the CMD $K_s$ vs. $J-K_s$ to select the RC stars. We generated 100 samples of randomly selected completeness corrected stellar lists and computed the extinction indices and $A_{1.61}$. We obtained that there is not any significant change of the measured values of the results for $\alpha_{JH}$, $\alpha_{HK_s}$, $A_{JH}$, and $A_{HK_s}$. Therefore, we conclude that completeness does not affect our results in any significant way within the selected stellar sample.

\section{The extinction index as a function of absolute extinction}

We studied the variability of the extinction index with the absolute extinction using two different approaches.

\subsection{Different GALACTICNUCLEUS regions}

\begin{figure}
   \includegraphics[width=\linewidth]{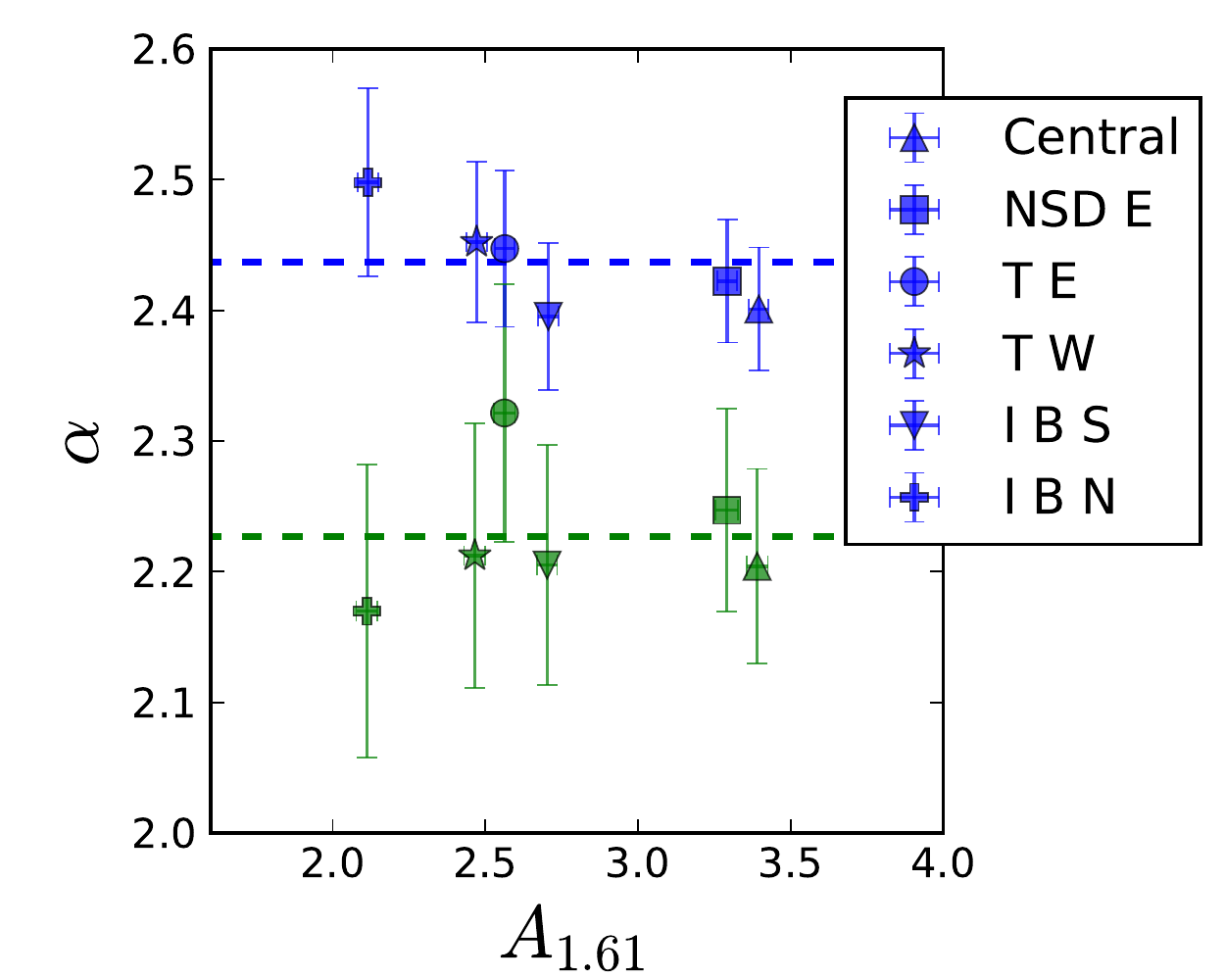}
   \caption{Extinction index variability as a function of the absolute extinction ($A_{1.61}$). Upper blue and lower green points indicate $\alpha_{JH}$ and $\alpha_{HK_s}$, respectively. The legend indicates the symbols corresponding to a given region of the GALACTICNUCLEUS survey. The uncertainties were computed considering the ZP systematics between regions. The green and blue dashed line indicate a flat profile to be compared with the extinction indices and their uncertainties.}

   \label{absolute_extinction_variability}
    \end{figure}

Using the grid method, we obtained $\alpha_{JH}$, $\alpha_{HK_s}$, $A_{1.61\_JH}$, and $A_{1.61\_HK_s}$ for six different regions of the GALACTICNUCLEUS survey (Table \ref{alpha_grid}). Given that the mean extinction varies significantly between them, we analysed the variability of the extinction indices with the absolute extinction obtained for each of the regions. Figure \ref{absolute_extinction_variability} shows the result. We computed the uncertainties assuming that the only significant relative difference between the studied regions corresponds to the possible ZP systematics.  We conclude that there is not any significant dependence of the extinction indices ($\alpha_{JH}$ and $\alpha_{HK_s}$) with the absolute extinction for the range of observed absolute extinctions within the obtained uncertainties. Computing the mean and the standard deviation of the six extinction indices, we ended up with $\alpha_{JH} = 2.44 \pm 0.03$ and $\alpha_{HK_s} = 2.23 \pm 0.05$.

\subsection{Grid method for different cuts in the CMD}

\begin{figure}
   \includegraphics[width=\linewidth]{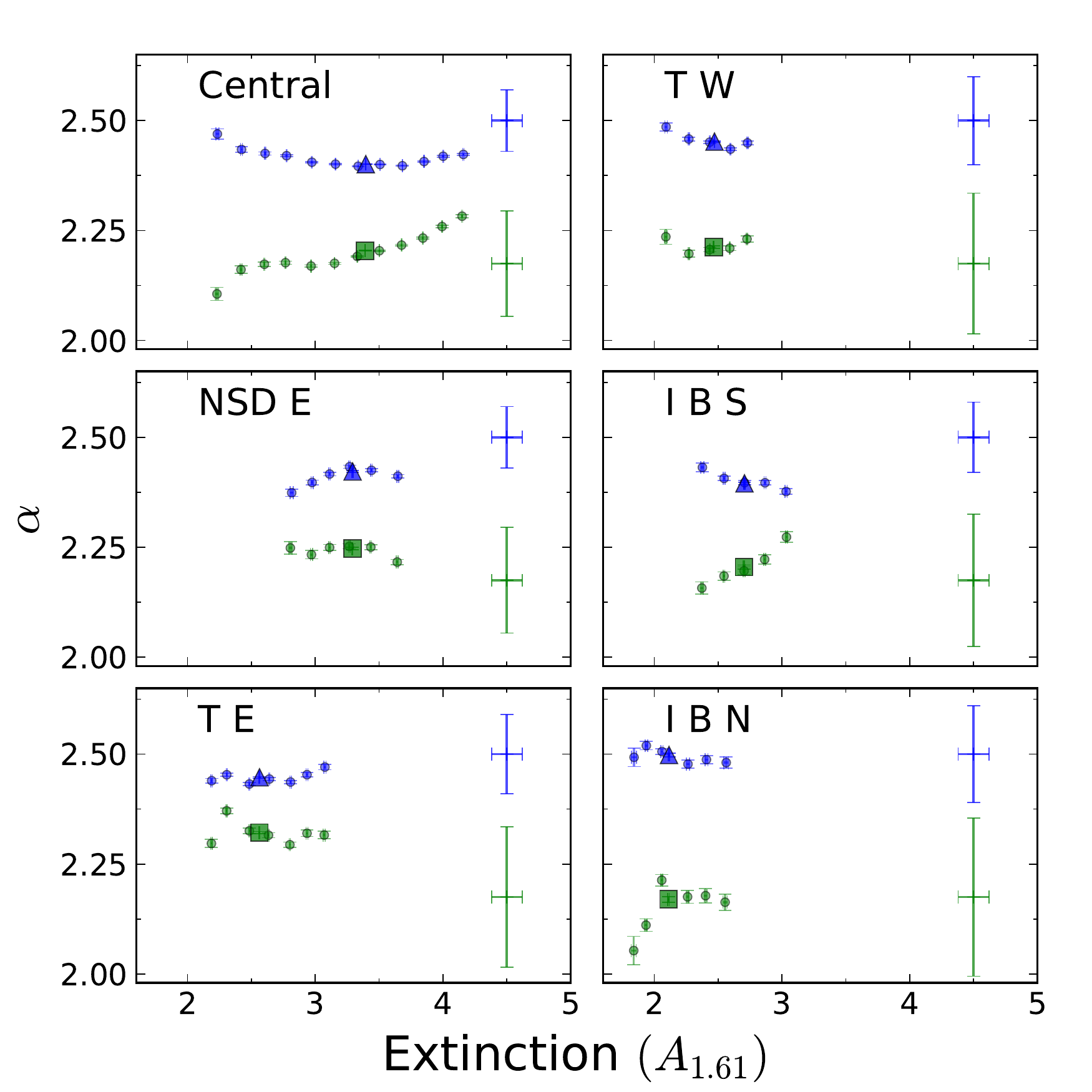}
   \caption{Extinction index variability with the absolute extinction ($A_{1.61}$) for six different regions of the GALACTICNUCLEUS survey (indicated for each panel). Upper blue points and lower green ones indicate the values of $\alpha_{JH}$ and $\alpha_{HK_s}$ for each absolute extinction $A_{1.61}$, including the statistical uncertainties. Big blue triangles and green squares indicate the average value per region obtained in Table \ref{alpha_grid}. The blue and green crosses indicate the systematic uncertainties due to the ZP computed for each region.}

   \label{absolute_extinction_6variability}
    \end{figure}

We also studied the variability of the extinction index with the absolute extinction within the same regions computing $\alpha_{JH}$ and $\alpha_{HK_s}$ using the grid method for different colour cuts in the CMD $K_s$ vs. $J-K_s$ as done in Sect. 4.3 of \citet{Nogueras-Lara:2019ac}. We obtained that $\alpha_{JH}$ is compatible with a constant extinction index, whereas $\alpha_{HK_s}$ shows some tendency to be larger at larger values of absolute extinction in the central, inner bulge South, and inner bulge North regions. Figure \ref{absolute_extinction_6variability} shows the results. In particular this difference is $\gtrsim 0.1$ for $\alpha_{HK_s}$ in the central region of the GALACTICNUCLEUS survey for a variation in $A_{1.61}\sim 2$\,mag. Nevertheless, this tendency does not appear for the nuclear stellar disk East, and the transition regions East and West. Moreover, the largest variations are observed for $\alpha_{HK_s}$ which is more sensitive to small changes in the parameters that affect the calculation (see Sect. \ref{grid}). Therefore, we believe that this tendency is not real.

On the other hand, and in particular for the central region where the variation is largest, the presence of a different foreground population (old and alpha enhanced from the inner bulge, e.g. \citet{Nogueras-Lara:2018ab}, might also explain slightly different values when computing the extinction index there. We concluded that the previously obtained mean values of $\alpha_{JH} = 2.44 \pm 0.03$ and $\alpha_{HK_s} = 2.23 \pm 0.05$, are more reliable and cover well the observed variations of the extinction indices within the uncertainties.

This approach of using colour cuts in the CMD $K_s$ vs. $J-K_s$  has several limitations that might introduce the tendency observed for the central, inner bulge South, and inner bulge North regions (see Fig. \ref{absolute_extinction_6variability}): 

\begin{itemize}

\item A vertical cut in the CMD $K_s$ vs. $J-K_s$ introduces selection effects on the stars used to compute the extinction indices. Namely, the intrinsic scatter in the plot and the partial degeneracy between the extinction and the extinction index impede to select a clean sample of stars corresponding to a given extinction. The shorter the bins, the larger this effect. Furthermore, the distribution of the extinction indices for a given cut is not fully Gaussian due to this effect. \\

\item The grid method using two bands to compute two unknowns ($\alpha$ and $A_{1.61}$) is similar to a geometric method in the CMD space and is dependent on the density of sources in the given RC feature. Therefore, cutting this feature into small bins introduces systematics effects that might produce a variation in the extinction index and absolute extinction with the number and density of RC stars for each bin. \\

%

\item We need to know the distance to the RC stars to compute $\alpha$ and $A_{1.61}$. The distance to the GC is a good estimate if we average over all the stars, but slightly different distances might become significant when considering colour cuts in the CMD. Moreover, the transparency varies between different regions of the NSD due to the presence or not of dusty clouds. Therefore, the contribution from the NSD's edge may vary, slightly changing the average distance to the observed stars. We estimated that a change in $\sim$200\,pc \citep[$\sim$\,diameter of the NSD, e.g.][]{Nishiyama:2013uq, gallego-cano2019} can produce a variation of $\alpha_{HK_s}$ around 0.05. This effect is less important for $\alpha_{JH}$, where it accounts for $\sim0.02$. This also applies to the previous effects and explains why we always observed a constant $\alpha_{JH}$ for all the analysed regions. The only way to reduce these effects is to select wide bins that contains a large number of stars in the RC feature, whose average distance is less biased, as we did in Sect. \ref{grid}.\\

\end{itemize}

\subsection{Effect on the RC feature in the CMD}

We also tested whether the variation in the extinction index as a function of absolute extinction would produce any significant effect on the slope of the RC features in the CMD $K_s$ vs. $H-K_s$. We assumed the  $\alpha_{HK_s}$ vs. $A_{1.61}$ values obtained for the central region in Fig. \ref{absolute_extinction_6variability} and computed the equivalent slope for each extinction bin, solving for $m$ in Eq. \ref{eq_slope}. We reconstructed the distribution of points in the $K_s$ vs. $H-K_s$ diagram following the changes of slope indicated by the $m$ values. Finally, we computed the slope of all the computed points using a jackknife algorithm to estimate the uncertainties (see Sect. 3.1 for further details). We obtained that the points are well fitted by a linear fit with a slope of $1.219\pm0.004$. On the other hand, the slope computed for the bright RC feature of the central region in Sect. \ref{slope_sec} was $1.217\pm0.010\pm0.015$, where the uncertainties refer to the statistical and the systematic ones, respectively. Therefore, given this variation in $\alpha_{HK}$, the behaviour of the RC slope is not significantly affected, so it is not possible to detect whether there is some variation (of this scale) in the extinction index with absolute extinction using the slope of the RC feature.

\section{Extinction index variability with the line of sight}

Figure \ref{absolute_extinction_variability} informs also about the variability of the extinction indices $\alpha_{JH}$ and $\alpha_{HK_s}$ as a function of the line of sight. We studied six different regions in the GC that are separated a maximum distance of $\sim 0.6^\circ$ ($\sim 90$ pc, see Fig. \ref{GNS}) and obtained that the extinction indices are compatible with being constant one within the uncertainties, as it was shown in the previous section.

\begin{figure*}
   \includegraphics[width=\linewidth]{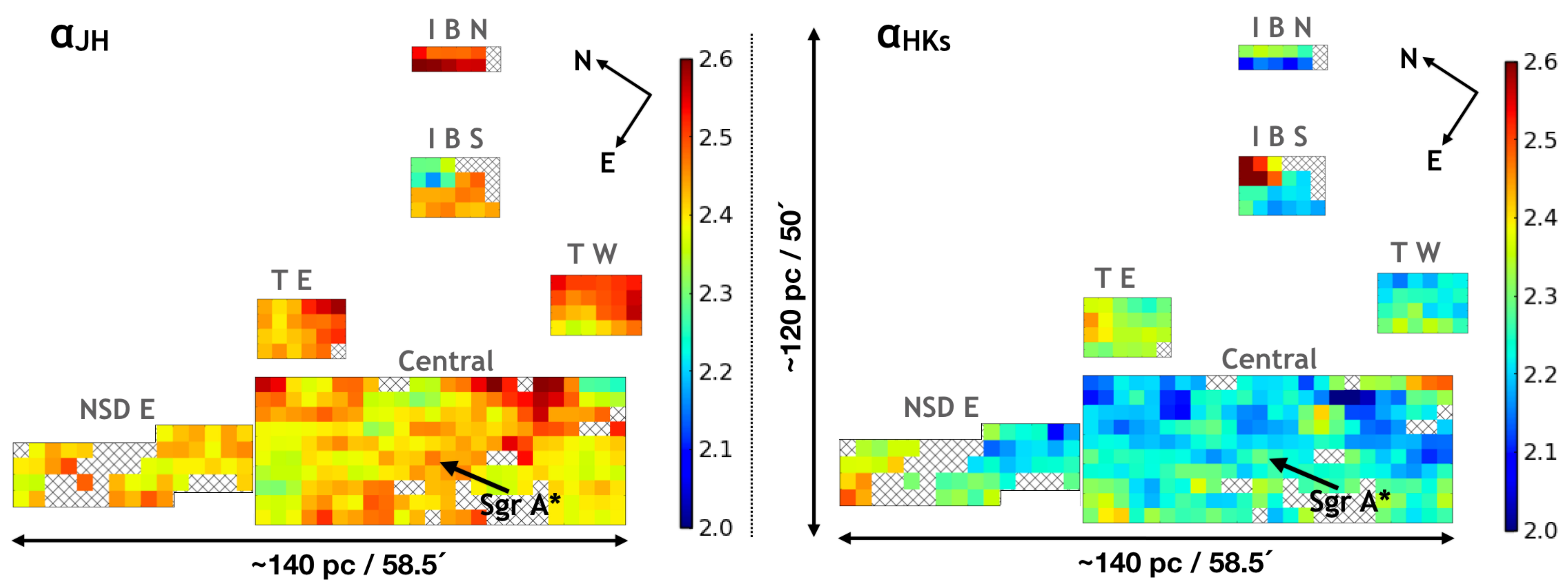}
   \caption{Line-of-sight distribution of the extinction indices $\alpha_{JH}$ (left panel) and $\alpha_{HK_s}$ (right panel). Cross-shaped pixels indicate regions where the number of stars is not enough to compute a value for the extinction index. The labels specify the analysed regions of the GALACTICNUCLEUS survey. The scale of the colour bar is the same for both panels. The position of Sgr A* and the physical scales are shown in the figure.}

   \label{line_of_sight}
    \end{figure*}

We also studied the variability of the extinction indices at shorter spatial scales. For this, we analysed all the regions in the GALACTICNUCLEUS survey with the exception of the NSD West region which has low data quality (see Sect. \ref{EIVW}). We used the extinction indices obtained for individual RC stars with the grid method, as previously explained. We created extinction index maps dividing the analysed regions into a grid of pixels of 1.5 arcmin$^2$. We computed the values for each pixel using a 3\,$\sigma$ clipping algorithm to reject outliers. We also excluded stars with photometric uncertainties larger than 0.05 mag in any single band and imposed a minimum number of 80 accepted stars to compute a pixel value \citep[for further details, see Sect. 4.1. of ][]{Nogueras-Lara:2019ac}. Figure \ref{line_of_sight} shows the maps obtained for $\alpha_{JH}$ and $\alpha_{HK_s}$. We chose the same colour scale for both maps to highlight the wavelength dependence of the extinction index. We estimated the statistical uncertainties for each pixel by means of the error of the mean of the $\alpha$ distributions (standard deviation /  (number of stars -1)$^{1/2}$). Considering the uncertainties for all the pixels, we obtained an average uncertainty of 0.01 for $\alpha_{JH}$, and 0.02 for $\alpha_{HK_s}$. The systematic uncertainties are irrelevant here because they affect all the values in the same direction. Nevertheless, since the GALACTICNUCLEUS survey is composed of 49 different pointings that were observed and photometrically calibrated in an independent way, the ZP systematics between different regions might cause a spurious variation. Considering the ZP systematic uncertainty of 0.04 for all the bands \citep{Nogueras-Lara:2019aa}, we estimated that the extinction indices might vary by $\delta\alpha_{JH} \sim 0.05$ and $\delta\alpha_{HK_s} \sim 0.08$. Therefore, the previous effects are sufficient to justify the variations between pixels observed in Fig.  \ref{line_of_sight}. Moreover, some pixels that appear to be different from the surrounding ones, normally correspond to the shape of each HAWK-I pointing and show a correlation between $\alpha_{JH}$ and $\alpha_{HK_s}$. This can be easily explained by a shift of the ZP for a given pointing in $H$ band, affecting both values, $\alpha_{JH}$ and $\alpha_{HK_s}$.

\begin{figure}
   \includegraphics[width=\linewidth]{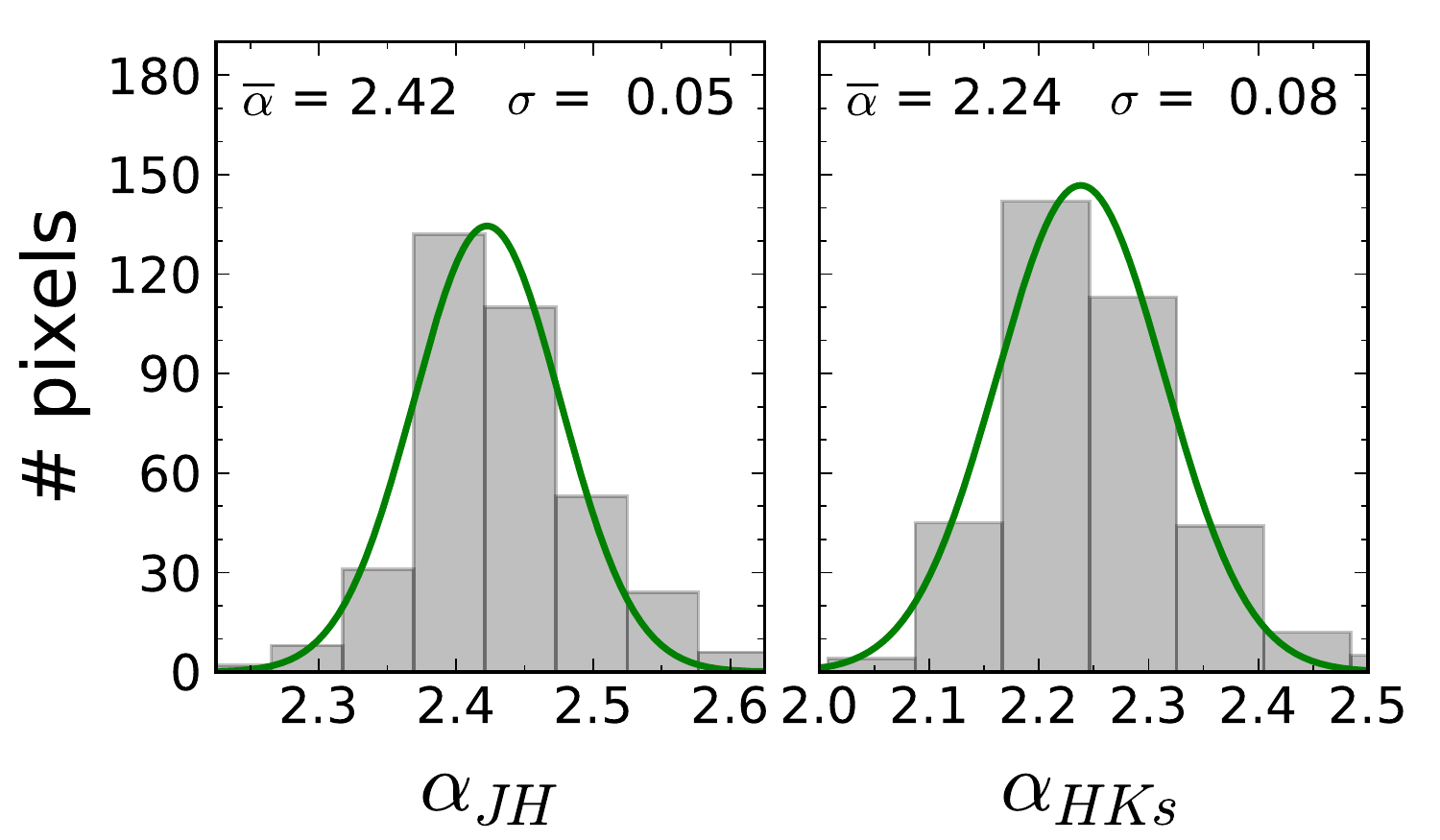}
   \caption{Histograms of the extinction indices $\alpha_{JH}$ (left panel) and $\alpha_{HK_s}$ (right panel) per pixels using all the regions shown in Fig. \ref{line_of_sight}. The green lines indicates a Gaussian fit whose mean and standard deviation values are specified on each panel.}

   \label{varia_map}
    \end{figure}

To further analyse the constancy of the extinction indices with the line-of-sight, Fig. \ref{varia_map} shows the distribution of the $\alpha_{JH}$ and $\alpha_{HK_s}$ obtained for all the pixels. We found that the data properly follow a Gaussian distribution with mean values of $\alpha_{JH} = 2.42\pm0.05$ and $\alpha_{HK_s} = 2.24\pm0.08$, where the uncertainties correspond to the standard deviation. These uncertainties are perfectly compatible with the systematics due to the photometric zero point computed previously. Hence, our results point towards non-varying extinction indices with the line-of-sight within the uncertainties.

\section{Final extinction index value}

We analysed the extinction curve in the NIR towards the GC and found that the extinction index depends significantly on the wavelength. On the other hand, it does not depend on the absolute extinction and on the line-of-sight within the estimated uncertainties. Now, we computed a final value for $\alpha_{JH}$ and $\alpha_{HK_s}$ combining the values obtained for the different methods and regions that we used. We took the average values computed using the slopes of the RC features in Table \ref{alpha_slope}, and all the individual values obtained using the grid method for the different regions of the GALACTICNUCLEUS survey (Table \ref{alpha_grid}). We ended up with a mean value of $\alpha_{JH} = 2.44\pm0.05$ and $\alpha_{HK_s} = 2.23 \pm 0.05$, where the uncertainties were quadratically propagated.

We also computed the ratios between absolute extinctions, $A_J/A_H$ and $A_H/A_{K_s}$ using the following equation:

\begin{equation}
A_{\lambda_1}/A_{\lambda_2} = (\lambda_1/\lambda_2)^{-\alpha} \hspace{0.5cm},
\end{equation}
\vspace{0.1cm}

\noindent where $\lambda_i$ indicates effective wavelength, $A_{\lambda_i}$ are the absolute extinctions for a given effective wavelength, and $\alpha$ is the corresponding extinction index. We computed the effective wavelengths considering the $A_{1.61}$ in Table \ref{alpha_grid} and obtained a consistent value of $A_J/A_H = 1.87\pm0.03$ and $A_{H}/A_{K_s} = 1.84\pm0.03$, where the uncertainties have been obtained quadratically propagating the associated uncertainties and considering the different values obtained using the different extinctions, $A_{1.61}$, associated to each of the regions of the GALACTICNUCLEUS survey.  

\section{Discussion and conclusion}

In this paper we analysed in detail the extinction curve towards the GC in the NIR applying independent methods based on RC stars. For the first time, we used the whole GALACTICNUCLEUS survey \citep{Nogueras-Lara:2019aa} and extended the study of the extinction to a large area of the NSD, the inner bulge, and the transition region between the GC and the inner bulge, covering a total area of $\sim 6000$ pc$^2$. For the analysis, we used more than $165,000$ stars in the RC feature detected in all three bands, $JHK_s$, of the GALACTICNUCLEUS survey, superseding by a factor $\sim3$ our previous study \citep{Nogueras-Lara:2019ac}.

We confirmed the wavelength dependence of the extinction curve in the NIR and derived values of $\alpha_{JH} = 2.44\pm0.05$ and $\alpha_{HK_s} = 2.23 \pm 0.05$, combining the results for different regions and methods. Thus, we estimated that the difference between both extinction indices is $\Delta\alpha = 0.21\pm0.07$, where the uncertainties were quadratically propagated. These values are compatible with the previous results ($\alpha_{JH} = 2.43\pm0.03$ and $\alpha_{HK_s} = 2.23 \pm 0.03$) obtained by \citet{Nogueras-Lara:2019ac}. The uncertainties of the new results are somewhat higher due to the larger size of the sample and the combination of values belonging to different regions and/or obtained by using independent methods. Our study confirms the evidence of a wavelength dependence also pointed out by \citet{Nogueras-Lara:2018aa,Hosek:2018aa}.

These new results might solve the apparent disagreement between the extinction indices obtained by independent groups that computed a single $\alpha$ for the NIR bands $JHK_s$ \citep[e.g.][]{Nishiyama:2006tx,Stead:2009uq,Gosling:2009kl,Fritz:2011fk,Alonso-Garcia:2017aa,Deno-Stelter:2020aa}. Moreover, using our results to correct the photometry from distance estimators, might lead to a change in the picture of the inner regions of the Milky Way \citep[e.g.][]{Matsunaga:2016aa}. Namely, a variation of $\sim 10-15\,\%$ in the extinction index, might produce a change of $\sim 0.3$ mag in absolute extinction that would lead to a biased distance calculation of around $\sim 1000$ pc for the GC distance, when using RC stars and the distance modulus for the estimation.

To compare our results with the extinction curve obtained by \citet{Hosek:2018aa}, we computed the extinction ratios $A_J/A_{K_s}$ and $A_H/A_{K_s}$ using the effective wavelengths that we have calculated for the central region of the GALACTICNUCLEUS survey (considering $A_{1.61}=3.40$ and the extinction indices that we derived). Table \ref{hosek_comp} shows the obtained results. The $A_J/A_{K_s}$ values agree within the uncertainties, while the $A_H/A_{K_s}$ ones are different. To analyse this disagreement, we computed the equivalent extinction indices $\alpha_{JH}$ and $\alpha_{HK_s}$ using the Table 5 from \citet{Hosek:2018aa} and the following equation:

\begin{equation}
\label{alphas_other}
\alpha_{\lambda_1\lambda_2} = \frac{log (A_{\lambda_1}/ A_{\lambda_2})}{log (\lambda_1/\lambda_2)} \hspace{0.5cm} ,
\end{equation}
\vspace{0.15cm}

\noindent where $A_{\lambda_i}$ refers to the extinction at a given wavelength $\lambda_i$. We obtained $\alpha_{JH} = 2.17\pm 0.07$ and $\alpha_{HK_s}=2.56\pm0.04$, where the uncertainties were quadratically propagated using the statistical uncertainties in Table \ref{hosek_comp}. In spite of being two different extinction indices, they show an inverted tendency with respect to our results. Comparing with previous work, we found that the vast majority of studies show a tendency that is compatible with our findings.  In particular, Fig. 7 from \citet{Nishiyama:2009oj} and Fig. 8 from \citet{Fritz:2011fk} summarise a variety of studies \citep[e.g.][]{Rieke:1985fq,Lutz:1999yf,Indebetouw:2005rp} that point towards a non-inverted tendency when moving towards redder wavelengths, in agreement to our results. Only the values obtained using hydrogen lines by \citet{Fritz:2011fk}  might present some evidence of inversion \citep[see Fig. 8 from][]{Fritz:2011fk}, but the final value is compatible with a single extinction index of $\alpha=2.11\pm0.06$.

\begin{table}
\caption{Results compared with \citet{Hosek:2018aa}.}
\label{hosek_comp} 
\begin{center}
\def\arraystretch{1.4}
\setlength{\tabcolsep}{3.8pt}

\begin{tabular}{ccc}
 &  & \tabularnewline
\hline 
\hline 
Parameter & This work &  \citet{Hosek:2018aa}\tabularnewline
\hline 
$A_J/A_{K_s}$  & 3.44 $\pm$ 0.08 & 3.52 $\pm$ 0.03\tabularnewline
$A_H/A_{K_s}$ & 1.84 $\pm$ 0.03 & 2.01 $\pm$ 0.05\tabularnewline
$\alpha_{JH}$ & 2.44 $\pm$ 0.05 & 2.17 $\pm$ 0.07\tabularnewline
$\alpha_{HK_s}$ & 2.23 $\pm$ 0.05 & 2.56 $\pm$ 0.04\tabularnewline
\hline 
 &  & \tabularnewline
\end{tabular}

\end{center}
\textbf{Notes.} We computed the values from \citet{Hosek:2018aa} using the code that they made publicly available (http://faun.rc.fas.harvard.edu/eschlafly/apored/extcurve\_s16.py) and computed the uncertainties quadratically propagating the obtained statistical uncertainties.

 \end{table}

Moreover, we further analysed this disagreement computing the extinction indices $\alpha_{JH}$ and $\alpha_{HK_s}$ using the Eq. \ref{alphas_other}, the extinction values from \citet{Nishiyama:2009oj} and  \citet{Schlafly:2016aa}, and the associated wavelength values that they give in their papers. We obtained $\alpha_{JH} = 2.10$ and  $\alpha_{HK_s} = 2.01$ and, $\alpha_{JH} = 2.27$ and $\alpha_{HK_s} = 2.24$, respectively. In the case of \citet{Schlafly:2016aa}, we used an $R_V = 3.3$, and adapted the extinction curve to our results using $A_{H}/A_{K_s}=1.84$  (instead of the value 1.55 that they used). We checked that, in spite of being consistent with a single extinction index, if there is some tendency, it would be in the direction of the wavelength dependence that we find. On the other hand, the extinction indices computed by \citet{Schodel:2010fk} for $\alpha_{HK_s}=2.21\pm0.24$ and $\alpha_{K_sL'}=1.34\pm0.29$, also support the tendency of a lower extinction index for redder NIR wavelengths. In this way, we believe that maybe a calibration problem might explain the inverted extinction indices tendency obtained by \citet{Hosek:2018aa}. Nevertheless, it is important to note that the majority of studies of the NIR extinction curve, including this one, use wide band filters for their analysis. This implies that they are limited by non-linear photometric effects \citep[for further details see][]{1980MNRAS.192..359J,2008BaltA..17..277S,Maiz-Apellaniz:2020aa}. In particular, the extinction affects a star in a different way depending on its spectral type and the effective wavelength (defined for wide-band filters), that varies depending on the absolute extinction and the type of star. In this work, we mainly used RC stars as reference stars to analyse the extinction, minimising the inclusion of other kind of stars that might contaminate our results. Moreover, we also defined the effective wavelength for the used RC stars independently for each of the regions, considering the appropriate average absolute extinctions as indicated in Table \ref{alpha_grid}. As a future goal, we aim at analysing the extinction curve using narrow band filters, that are less affected by these issues and will help to determine the precise behaviour of the extinction curve.

We showed that the evidence of the $\alpha_{HK_s}$ dependence as a function of the absolute extinction in \citet{Nogueras-Lara:2019ac} was not a real effect and it is due to systematics in the methodology and some degeneracies, such as the mixture of stars that might be located at different distances in the NSD. Thus, after analysing GALACTICNUCLEUS regions whose extinction is significantly different, we concluded that the extinction index does not show any significant variation with the absolute extinction within the estimated uncertainties.

Finally, we also found that the calculated extinction indices, $\alpha_{JH}$ and $\alpha_{HK_s}$, do not depend on the line-of-sight and can be assumed constant within the given uncertainties and the analysed regions, in agreement with previous work \citep{Nogueras-Lara:2019aa, Deno-Stelter:2020aa}. For this analysis, we considered all the fields observed in the GALACTICNUCLEUS survey, but the NSD W, whose low quality impeded this study. In this way, we covered regions that are within one square degree without finding any significant line-of-sight variation.

  \begin{acknowledgements}
      The research leading to these results has received funding from
      the European Research Council under the European Union's Seventh
      Framework Programme (FP7/2007-2013) / ERC grant agreement
      n$^{\circ}$ [614922]. This work is based on observations made with ESO
      Telescopes at the La Silla Paranal Observatory under programmes
      IDs 195.B-0283 and 091.B-0418. We thank the staff of
      ESO for their great efforts and helpfulness.

F. N.-L. and N. N. gratefully acknowledge support by Sonderforschungsbereich SFB 881 ‘The Milky Way System’ (subproject B8) of the German Research Foundation (DFG). R. S., A. T. G.-C., E. G.-C., and B. S. acknowledge financial support from the State
Agency for Research of the Spanish MCIU through the "Center of Excellence Severo
Ochoa" award for the Instituto de Astrof\'isica de Andaluc\'ia (SEV-2017-0709). A. T. G.-C., B. S., and R. S.  acknowledge financial support from national project
PGC2018-095049-B-C21 (MCIU/AEI/FEDER, UE). F. N. acknowledges financial support through Spanish grant ESP2017-86582-C4-1-R (MINECO/FEDER) and from the  Spanish State Research Agency (AEI) through the Unidad de Excelencia “María de Maeztu” -Centro de Astrobiología (CSIC-INTA) project  No. MDM-2017-0737.
\end{acknowledgements}

\bibliography{../BibGC.bib}

\end{document}